\documentclass{article}
\usepackage{epsfig}
\usepackage{amsmath}

\date{\today}

\newcommand{\vect}[1]{\stackrel{\rightarrow}{#1}}
 
\newcommand{\insertplot}[5]{\begin{figure}
 \hfill\hbox to 0.05in{\vbox to #5in{\vfill
 \inputplot{#1}{#4}{#5}}\hfill}
 \hfill\vspace{-.1in}
 \caption{#2}\label{#3}
 \end{figure}}
 \newcommand{\inputplot}[3]{
 \special{ps: plotfile #1}
}
\newcounter{fig}   

\usepackage{epsfig} 
\usepackage{amsfonts}
\usepackage{graphicx}
\usepackage{epsfig} 

\usepackage{a4}

\usepackage{amsmath}

\date{\today}

\begin{document}
\title{A note on Klein-Gordon equation in a  generalized 
\\
Kaluza-Klein monopole background}
 \author{{\large {\bf Eugen Radu}}$^{\dagger}$ and
{\large {\bf Mihai Visinescu}}$^{\ddagger}$  
\\ \\
$^{\dagger}${\small Department of
Mathematical Physics,  National University of Ireland, Maynooth, Ireland}
\\
$^{\ddagger}$
{\small   Department of Theoretical Physics, National
Institute for Physics and Nuclear Engineering,}
\\ 
{\small  P.O.Box M.G.-6,
Magurele, Bucharest, Romania} 
 }
\maketitle
 
\begin{abstract} 
We investigate solutions of the Klein-Gordon equation in a 
class  of five dimensional geometries
presenting the same symmetries and asymptotic structure
as the Gross-Perry-Sorkin monopole solution.
Apart from globally regular metrics, we consider also squashed Kaluza-Klein 
black holes backgrounds. 
\end{abstract}

\section{Introduction}
Recently, there has been an increasing interest in the solutions of 
Einstein equations involving more than four dimensions.
Such configurations are important if one supposes 
the existence of extra dimensions in the universe, 
which are likely to be compact and described by a Kaluza-Klein (KK) theory.

A particularly interesting case are the $d=5$ 
asymptotically locally flat configurations, approaching a twisted
$S^1$ bundle over a four dimensional Minkowski spacetime.
The best known example with this structure is the  
Gross-Perry-Sorkin (GPS)  monopole solution
\cite{Gross:1983hb,Sorkin:1983ns}.
Various generalizations of this
solution  with similar asymptotics have been discussed in the literature,
representing both globally regular and black hole configurations.

The study of classical fields in these backgrounds is a reasonable task as
 a first step towards their quantization.
Also, the GPS background provides one of the few examples available in the 
literature where 
it is possible to solve in closed form various field equations on a curved background.
The purpose of this work is to study the solutions of the scalar wave equation 
in a generic $d=5$ asymptotically locally flat spacetime,
presenting the same symmetries  as the GPS monopole.
We find that in the globally regular case, the properties of the
solutions are rather background independent. 

\section{The general framework}
\subsection{Generalized Kaluza-Klein monopole space-times}

We consider generalized $(4+1)$-dimensional KK monopole
space-times,
with the  metric given by line elements of the form
\begin{eqnarray}
\label{(met)}
ds^{2}=g_{\mu\nu}dx^{\mu}dx^{\nu} 
=-h dt^{2}+f \big(dr^{2}+ r^{2}(d\theta^{2}+ 
\sin^{2}\theta\, d\phi^{2})\big)+ g (d\chi+\cos\theta\, d\phi)^{2}\,,  
\end{eqnarray}
written in spherical coordinates, $r, \theta, \phi$, commonly related 
with 
the Cartesian ones  and $\chi$ defined as usual, while $t$ is the time 
coordinate. 
The coordinates $\theta$ and $\phi$ cover the sphere $S^{2}$ while 
$\chi\in D_{\chi}=[0,4\pi)$. 
We  suppose that the functions $f,g$ and 
$h$ depend only on the radial coordinate $r$. In addition 
we assume that on the radial domain of 
the local chart, $D_{r}$, these  functions  are positive definite. 
Obviously, the whole physical space domain of this chart is  
$D=D_{r}\times S^{3}$.
 
By construction, the spaces with the metric (\ref{(met)})  have 
five Killing vectors. The corresponding constants of 
motion consist of a  conserved 
quantity 
for the cyclic variable $\chi$, 
$q = g(r) (\dot\chi + \cos\theta \dot\varphi)$,~
the energy,
and the angular momentum vector
$\vec{J}=\vec{x}\times\vec{p}\,+\,q\,\frac{\vec{x}}{r}$, with
$\vec{p} = f(r)\dot{\vec{x}} $.

The line element (\ref{(met)}) may describe two different types of 
configurations.
The first one corresponds to (topologically nontrivial) globally regular 
spacetimes, 
with $h>0$ for the whole range of $r$.
The structure of these solutions is determined by the
dimension of the fixed point set of 
 the Killing vector
$\partial/\partial \chi$.
In general, this $U(1)$ isometry can have a zero-dimensional
fixed point set (referred to as a "nut" solution) or a two-dimensional
point set in the four dimensional
Euclidean space (correspondingly referred to as "bolt" solution).
The standard KK  monopole solution
\cite{Gross:1983hb,Sorkin:1983ns}
corresponds to the first case and has  
\begin{equation}
\label{nut}
h=1\,,\quad f=1+\frac{4m}{r}\,,\quad g=\frac{16 m^2}{f}\,,  
\end{equation}
with $m$ a constant. 
The $r=0$ corresponds here to the origin of the coordinate system in $R^4$.
When taking instead  the product of the $d=4$ Euclidean Taub-bolt solution \cite{Page:1979aj}
with the real line, the metric functions are
\begin{eqnarray}
\label{bolt}
h=1\,,\quad f=1+\frac{m}{4r}+\frac{9m^2}{64 r^2}\,,\quad 
g=\frac{4 m^2}{f} (1-\frac{9m^2}{64 r^2})^2 ~,
\end{eqnarray}
with $r\geq r_b=3m/8$.

Apart from these configurations, there are also squashed KK black holes,
with an event horizon located at $r=r_h$ with $h(r_h)=0$ and $f,g$ 
nonvanishing, while $f,g,h$ stay positive
for any $r\geq r_h$ (see $e.g.$ \cite{Ishihara:2005dp}-\cite{Brihaye:2006ws}). 
For example, the line element of the 
Einstein-Maxwell squashed  black hole found by 
Ishihara and Matsuno, expressed in the coordinate system used for the
metric ansatz  (\ref{(met)}), reads
\begin{eqnarray}
\nonumber
f(r)&=&\frac{1}{16 r^4}
(r^2+\frac{1}{2r}(2r_0+r_m+r_p)+\frac{1}{16}(r_p-r_m)^2)
((4r+r_p)^2-2r_m(r_p-4r)+r_m^2)~,
\\
\label{IM}
g(r)&=&\frac{(r_0+r_m)(r_0+r_p)((4r+r_p)^2-2r_m(r_p-4r)+r_m^2)}
{16r^2+8r (2r_0+r_m+r_p)+(r_m-r_p)^2},
\\
\nonumber
h(r)&=&\frac{((r_m-r_p)^2-16r^2)^2}{(4r+r_p)^2-2r_m(r_p-4r)+r_m^2}~,
\end{eqnarray}
where $r_0$, $r_m$ and $r_p$ are real parameters
related to the mass, electric charge and the size of the extra dimension, and 
$r_h=(r_p-r_m)/4$.
For completeness, we present also the U(1) potential expression 
(the Lagrangian in this case is $L=R/(16 \pi G)-F^2/4)$, with $F=dA$)
\begin{equation}
\label{IM-U1} 
\nonumber
A=4\sqrt{\frac{3}{\pi G}}\frac{r\sqrt{r_m r_p}}{(4r+r_m)^2+
8 rr_p-2 r_mr_p+r_p^2}dt~,
\end{equation}
(the choice $r_m=0$ gives the vacuum
black version of the GPS monopole presented in \cite{Chen:1999rv}). 
No similar closed form expression can be written for a more complicated 
matter content ($e.g.$ a nonabelian field).

In the general case, the expression of the metric 
function $f,~g$ and $h$ is usually fixed by the matter content
of the theory and the boundary conditions we choose.
We shall suppose, however, that, similar to the vacuum cases (\ref{nut}), 
(\ref{bolt}),
the spacelike infinity is always a squashed sphere or $S^1$ bundle over $S^2$.
Thus, for large values of $r$, one finds 
\begin{eqnarray}
\label{inf-exp}
f(r)=1+\frac{\tilde{f_1}}{r}+O(\frac{1}{r^2}),~~
h(r)=1-\frac{2 M}{r}+O(\frac{1}{r^2}),~~
g(r)=16 m^2(1+\frac{\tilde{g_1}}{r})+O(\frac{1}{r^2}),
\end{eqnarray}
where $\tilde{f_1}, M,~\tilde{g_1}$ and $m$ are real constant
(with $m$ and $M$ fixing the size of the extra dimension and total mass-energy 
\cite{Mann:2005cx}).

\subsection{The Klein-Gordon equation}

The equation governing the behaviour of a massless, minimally coupled
scalar field is (in natural units with $\hbar=c=1$), 
\begin{equation}
\label{(kg)}
\frac{1}{\sqrt{-g}}\frac{\partial }{\partial x^\mu}(\sqrt{-g}g^{\mu \nu}
\frac{\partial \psi}{\partial x^\mu})=0~,
\end{equation} 
and has  particular solutions 
\begin{equation}\label{(sol)}
\psi(x)=U_{E}({\bf x},\chi)e^{-iEt}\,, 
\end{equation}
of given frequency $E$.
The separation of variables in Eq.(\ref{(kg)}) can be done  if one takes
\begin{eqnarray}
\label{U}
U_{E}({\bf x},\chi)= 
{\cal P}(r)\,Y_{l,m}^{q}(\theta, \phi, \chi)~,
\end{eqnarray}
where $Y_{l,m}^{q}(\theta, \phi, \chi)$ are the $SO(3)\otimes U(1)$ harmonics
 \cite{Cotaescu:1999gu}.
These harmonics are eigenfunctions of the operators 
\begin{eqnarray}
{\vect{L}\,}^{2}=-\frac{1}{\sin\theta}\,\partial_{\theta}\,(\sin\theta\,
\partial_{\theta})-
\frac{1}{\sin^{2}\theta} (\partial_{\phi}^{2}+\partial_{\chi}^{2}-
2\cos\theta\,\partial_{\phi}\,\partial_{\chi}),~~
L_3=\partial_{\phi},~~Q=\partial_{\chi},~~
\nonumber
\end{eqnarray}
with 
\begin{eqnarray}
\nonumber
{\vect{L}\,}^{2}Y_{l,m}^{q}=l(l+1)\,Y_{l,m}^{q}\,, 
~~
L_{3}Y_{l,m}^{q}=m\,Y_{l,m}^{q}\,,
\label{(l3)}
~~
\nonumber
QY_{l,m}^{q}=q\,Y_{l,m}^{q}\,,
\end{eqnarray}
and satisfy the orthonormalization condition 
\begin{eqnarray}
&&\left<Y_{l,m}^{q},Y_{l',m'}^{q'}\right>=
\int_{S^2}d(\cos\theta)d\phi\,\int_{0}^{4\pi}d\chi\,
{Y_{l,m}^{q}(\theta, \phi, \chi)}^{*}\,
Y_{l',m'}^{q'}(\theta, \phi, \chi) 
=\delta_{l,l'}\delta_{m,m'}
\delta_{q,q'}\,,\label{(spy)}
\end{eqnarray}  
Notice that the boundary conditions on $S^{2}\times D_{\chi}$ require $l$ and 
$m$ to be integer numbers while $q=0,\pm 1/2,\pm 1,...$ \cite{CFH}. The form 
of these harmonics and a discussion of their properties
is given in \cite{Cotaescu:1999gu}.

Thus, one finds that the function ${\cal P}(r)$ in (\ref{U}) is a solution 
of the equation
\begin{eqnarray}
\frac{1}{r^2 f \sqrt{fgh}}\frac{d}{dr}(r^2\sqrt{fgh}\frac{d{\cal  P}}{dr})
-\frac{1}{r^2f}(l(l+1)-q^2){\cal P}
-\frac{q^2}{g}{\cal P}+\frac{E^2}{h}{\cal P}=0.
\end{eqnarray}
This radial equation
can be transformed to a standard Schr\"odinger form
be defining a "tortoise" radial variable $dr_*=\sqrt{f/h}dr$ and taking 
${\cal P}(r)= F(r)/F_1(r)$,
with 
$F_1(r)=(rf^{1/2}g^{1/4})$.
 One finds
\begin{eqnarray}
\label{n2}
\nonumber
 -\frac{d^2F}{dr_*^2}+V(r)F=E^2 F~,
\end{eqnarray}
with a potential
\begin{eqnarray}
\label{n3}
\nonumber
V(r)=\frac{h}{r^2f}(l(l+1)-q^2)+\frac{q^2h}{g}-
\frac{1}{F_1^3}\sqrt{\frac{h}{f}}(r^2\sqrt{fgh}F_1')'~,
\end{eqnarray}
(where a prime denotes the derivative w.r.t. $r$).

However, in practice we found more convenient to take
\begin{equation}
U_{E}({\bf x},\chi)=\frac{1}{\rho(r)} 
R_{E,l}^{q}(r)\,Y_{l,m}^{q}(\theta, 
\phi, \chi)~,
\end{equation}
where a suitable choice for the function $\rho$ is
 \begin{equation}
\label{ro}
 \rho=r|fgh|^{1/4}\,. 
 \end{equation}
Then, after a few manipulation, we find the radial equation
\begin{eqnarray}
\label{(radeq)}
&&\left[-\frac{d^2}{dr^2}+\frac{1}{r^{2}}[l(l+1)-q^{2}]+
\frac{f}{g} q^{2}+\frac{1}{\rho}\frac{d^{2}\rho}{
dr^2}\right] R_{E,l}^{q}(r)
= E^2\frac{f}{h}R_{E,l}^{q}(r)~.
\end{eqnarray}
The radial function $R_{E,l}^{q}$ is normalized according to the radial scalar product
\begin{equation}\label{(scprod)}
\left<R_{E,l}^{q},R_{E',l}^{q}\right>=\int_{D_{r}}dr \,\frac{f}{h} 
(R_{E,l}^{q})^{*} R_{E',l}^{q}\,~,
\end{equation} 
resulted from the fact that  the $SO(3)\otimes U(1)$  harmonics, 
$Y_{l,m}^{q}$, 
are normalized to unity with respect to their own scalar product. 
The radial equation (\ref{(radeq)}) is similar to those of the non-relativistic 
quantum 
mechanics apart the term
${\rho''}/{\rho}~.$
 
Unfortunately, there is only one solution of the equation (\ref{(radeq)})
known in closed form, corresponding to a 
GPS monopole background \cite{Cotaescu:1999gu}\footnote{As usual
in metric backgrounds with $g_{tt}=-1$, the expression of the
scalar field $\Psi$
can be read from the solutions of the Schr\" odinger equation in 
\cite{Cotaescu:1999gu} 
by replacing $E \to E^2/2$ in the relations there.}.
However, one can analyze the properties of the general solutions  
 by using a combination of analytical and numerical
methods, which is enough for most purposes.
For example, the leading order
asymptotic expansion of the radial function $R(r)$ is
shared by all solutions. 
Taking into account the expressions (\ref{inf-exp})
of the metric functions, 
one finds that the equation (\ref{(radeq)})
presents an effective mass term in its asymptotic expansion,
$R''+(E^2 -\frac{q^2}{4m^2})R=0$.
Thus, for $E>|q|/2m$, the radial 
function has an oscillatory behaviour at infinity
\begin{eqnarray}
\label{asi1}
R_{E,l}^{q}(r)\sim e^{-i\sqrt{E^2-q^2/(4m^2)} r}+
s(E)e^{i\sqrt{E^2-q^2/(4m^2)} r}~.
\end{eqnarray}
For $E<|q|/2m$, the radial function decays asymptotically according to
\begin{eqnarray}
\label{asi2}
R_{E,l}^{q}(r)\sim e^{-\sqrt{ q^2/(4m^2)-E^2} r}~.
\end{eqnarray}
 
\section{Solutions in a globally regular  background}
\subsection{The Iwai-Katayama background}
 
An interesting case of globally regular backgrounds
are the metrics proposed by Iwai and Katayama (IK)
\cite{IK1,IK2,IK3}, with
\begin{equation}
\label{gtnut}
h=1\,,\quad f(r)=\frac{a}{r}+b\,,\quad 
g(r)=\frac{ar+br^{2}}{1+cr+dr^{2}}\,,
\end{equation}
where $a, b, c, d$ are constants\footnote{These configurations are of interest
mainly because they admit a Kepler-type symmetry. Other generalized Taub-NUT 
metrics could have a self-dual Weyl curvature tensor, be conformally 
flat \cite{IK1},
or produce complete Einstein self-dual metrics on $4$-balls \cite{YM}, etc.}.
If one takes these constants 
$c= {2 b}/{a}, ~d =  {b^2}/{a^2}$
the four dimensional generalized Taub-NUT part of the metric 
(\ref{(met)})
becomes the original
Euclidean Taub-NUT metric up to a constant factor.

The remarkable result of Iwai and Katayama is that the generalized 
Taub-NUT 
space (\ref{gtnut})  admits a hidden symmetry represented by a 
conserved vector, 
quadratic in 
$4$-velocities, analogous to the Runge-Lenz vector of the following 
form
\begin{equation}\label{rl}
\nonumber
\vec{K} = \vec{p} \times \vec{J} + \kappa \frac{\vec{x}}{r}\,.
\end{equation}
The constant $\kappa$ involved in the Runge-Lenz vector (\ref{rl}) is
$ \kappa = - a\,E + \frac{1}{2} c\,q^2 $
where the conserved energy $E$ is
$E =  {\vec{p}^{~2}}/{2 f(r)} +  {q^2}/{2 g(r)} \,.$
The components $K_i=k^{\mu\nu}_i p_{\mu}p_{\nu}$ of the vector 
$\vec{K}$ 
(\ref{rl}) involve three St\"ackel-Killing tensors $k^{\mu\nu}_i,~ i = 1,2,3$.

By rescaling the radial coordinate $r$, one can always set
$b=1$ in the general metric functions (\ref{gtnut}).
It is also convenient to define $c=Ca, ~d=Da$, the line element 
(\ref{(met)}) becoming
\begin{eqnarray}
\label{(met1)}
ds^{2}=
-dt^{2}+(1+\frac{a}{r})(dr^{2}+ r^{2}(d\theta^{2}+ 
\sin^{2}\theta\, d\phi^{2}))+  
\frac{a^2(1+\frac{a}{r})}{D+\frac{2Ca}{r}
+\frac{a^2}{r^2}}(d\chi+\cos\theta\, d\phi)^{2}\,,  
\end{eqnarray}
We suppose that the coefficients $C,~D$ are restricted such that
the function $D+\frac{2Ca}{r}
+\frac{a^2}{r^2}$ takes only positive values on $D_r$.
Also, to simplify the general picture we shall suppose $a>0$.

\subsubsection{A perturbative solution}
The  radial equation (\ref{(radeq)}) can be written as
$\hat O R(r)=0,$
with the operator
\begin{eqnarray}
\label{eqn}
\nonumber 
\hat O =  -\frac{d^2}{dr^2}+\frac{1}{r^{2}}[l(l+1)-q^{2}]+
\frac{f}{g} q^{2}+\frac{1}{\rho}\frac{d^{2}\rho}{
dr^2}-E^2\frac{f}{h}~.
\end{eqnarray}
%
The choice $C=D=1$ in the generic line element (\ref{(met1)}) 
corresponds to a standard GPS monopole background.
This suggest a perturbative approach for the  equation
(\ref{(radeq)}), by taking
the expansion
\begin{eqnarray}
\label{ans1}
C= \sum_{k=0}^{\infty} c_k \varepsilon^k,~~D=\sum_{k=0}^{\infty} d_k 
\varepsilon^k,~~
R(r)= \sum_{k=0}^{\infty} R_k(r) \varepsilon^k,
\end{eqnarray}
with $c_0=d_0=1$ and $\varepsilon$ a small parameter.
This implies the following decomposition of the operator $\hat O$  
\begin{eqnarray}
\hat O=\hat O_{(0)}+\sum_{k>0}\hat O_{(k)} \varepsilon^k
\end{eqnarray}
with 
\begin{eqnarray}
\nonumber
&\hat O_{(0)}=   -\frac{d^2}{dr^2}+\frac{1}{r^{2}}[l(l+1)-q^{2}]+
\frac{(1+a/r)^2}{a^2} q^{2} - E^2(1+\frac{a}{r}),
\\
\nonumber
&\hat O_{(k)}=\frac{1}{a^2}(d_k+\frac{2c_ka}{r})+
\big(\frac{\rho''}{\rho}\big)_{(k)},~~{\rm with}~k>1,
\end{eqnarray}
where the first terms in the general expression of 
$ ( \rho''/\rho)_{(k)}$ are
\begin{eqnarray}
\nonumber
&\Big(\frac{\rho''}{\rho}\Big)_{(1)}=-
\frac{a^2}{2r(a+r)^4}(2ac_1+(3d_1-4c_1)r),~~~
\Big(\frac{\rho''}{\rho}\Big)_{(2)}=
\frac{a^2}{4r(a+r)^6}\big(-4a^3c_2
\\
\nonumber
&+a^2(13c_1^2-6d_2)r
-2a(6d_2-6c_2+13c_1(c_1-d_1))r^2
+(5c_1^2+8c_2-18c_1d_1+11d_1^2-6d_2)r^3\big).
\end{eqnarray}
One can see that $R_0(r)$ solves the equation (\ref{(radeq)})
for a standard GPS monopole background, $\hat O_{(0)}R_0(r)=0$.
The solutions of this equation have been discussed in \cite{Cotaescu:1999gu}.
The $k-$th order equation reads
\begin{eqnarray}
\sum_{i=0}^{k}\hat O_{(i)}R_{k-i}=0,~~{\rm
or~}~~ 
\label{geneq}
\hat O_{(0)} R_k=
S_k(r)=-(\hat O_{(1)} R_{k-1}(r)+\dots +\hat O_{(k)} R_{0}(r))~.
\end{eqnarray}
The general solution  of the above equation compatible with
the required boundary conditions reads \cite{morse}
\begin{eqnarray}
R_k(r)=R_0(r)+\int_{D_r}  G(r,r')S_k(r')dr'~,
\end{eqnarray}
where $G(r,r')$ is the Green function associated with 
radial equation (\ref{(radeq)})
 for a GPS monopole background,
  $\hat O_{(0)}G(r,r')=-\delta(r,r')$, $i.e.$ 
\begin{eqnarray}
G(r,r')=\frac{f_{>} (r_{>} )f_{<} (r_{<} )}{W(f_{>} ,f_{<} )}~,
\end{eqnarray}
where
\begin{eqnarray}
\nonumber
&&f_1(r)=e^{-\sqrt{q^2- a^2E^2}r/a}r^{1+l}
L(-1-l+\frac{-q^2+a^2E^2/2}{\sqrt{q^2- a^2E^2}},1+2l,
\frac{2\sqrt{q^2- a^2E^2/2}}{a},r),
\\
\nonumber
&&f_2(r)=e^{-\sqrt{q^2- a^2E^2}r/a}r^{1+l}
U(1+l+\frac{q^2-a^2E^2/2}{\sqrt{q^2- a^2E^2}},2(1+l),
\frac{2\sqrt{q^2-2 a^2E^2}}{a},r),
\end{eqnarray}
are the independent solutions to the equation 
$\hat O_{(0)}f(r)=0$, with $U$ and $L$ the confluent 
hypergeometric functions and 
the generalized Laguerre polynomial, respectively \cite{morse}.
 $f_{>} (r_{>} )=f_{2} (r_{>} )$
 satisfies the boundary condition of finiteness at 
infinity
and $f_{<} (r_{<} )=f_{1} (r_{<} )$
 is similar finite as $r$ goes to zero; $W$ is the wronskian of $f_>$ 
and $f_<$ (with $r_{<}={\rm min}(r,r'),~r_{>}={\rm max}(r,r')$).

\subsubsection{Numerical solutions}
One can also look  for nonperturbative solutions of the equation 
(\ref{(radeq)}). Here we need the asymptotic expansion of the solution 
near the origin and at infinity. These asymptotics are very similar to 
the case of solutions in the background (\ref{nut}).
A systematic analysis gives
\begin{eqnarray}
\label{as1}
 R(r) \sim r^s,~{\rm as~~}r \to 0,
\end{eqnarray}
where the parameter $s$ is a solution of the equation
$s(s-1)=l(l+1)$.
The modes with $s=l+1$, called in \cite{Cotaescu:1999gu}
$\it{regular~modes}$ are similar to the those of the usual 
nonrelativistic case.
Apart from these, there are also $\it{irregular~modes}$ with $s=-l$,
which are relevant in the $a<0$ case (not considered here).

As $r\to \infty$ one finds the following asymptotic form
\begin{eqnarray}
\label{as2}
 R(r) \sim e^{-\lambda r},~~{\rm with~~}\lambda=\sqrt{\frac{Dq^2}{a^2}- E^2}~,
\end{eqnarray}
the general solution being on the form
\begin{eqnarray}
\label{as3}
 R(r) = r^s e^{-\lambda r} P(r)~.
\end{eqnarray}
%
Taking $s=l+1$, one finds that satisfies $P(r)$ the equation
\begin{eqnarray}
\label{p1}
-P''+(-\frac{2(l+1)}{r}+2\sqrt{\frac{Dq^2}{a^2}- E^2})P'(r)
+K(r)P(r)=0
\end{eqnarray}
where $K(r)$ has a complicated expression.
However, as $r\to 0$ one can write 
\begin{eqnarray}
\label{f1}
K(r)= \frac{1}{ar}(1- a^2E^2+C(2q^2-1)+2a\sqrt{Dq^2-a^2E^2}(l+1))~~~{~~}
\\
\nonumber
-\frac{5+(2-13C)C+6D}{4a^2}
+
\frac{r}{2a^3}\left(3-D+C(1+2(1-9C)C+13D)\right)  
+O(r^2),
\end{eqnarray}
 its asymptotic expansion as $r \to \infty$ being
\begin{eqnarray}
\label{f2}
K(r)=
 \frac{2}{ar}\big(-a^2E^2/2+Cq^2+\sqrt{Dq^2- a^2E^2}(l+1)\big)+O(1/r^3)~ .
\end{eqnarray}
One can see from  (\ref{as2}) that, in agreement with the relation (\ref{asi1}), for
$E^2>Dq^2/{a^2}$ there is only a continuous energy spectrum,
the levels of which are infinite degenerate.

The case $E^2<Dq^2/{a^2}$ is more involved, since
one finds also a discrete sector of the spectrum.
The arguments here are similar to the case $C=D=1$ \cite{Cotaescu:1999gu}.
Following the standard approach \cite{morse}, we suppose that $P(r)$ admits a 
power series expansion
\begin{eqnarray}
\label{series}
P(r)=\sum_k p_k r^k,
\end{eqnarray}
%
\begin{figure}[ht!]
\parbox{\textwidth}
{\centerline{
\mbox{
\epsfysize=18.0cm
\includegraphics[width=92mm,angle=0,keepaspectratio]{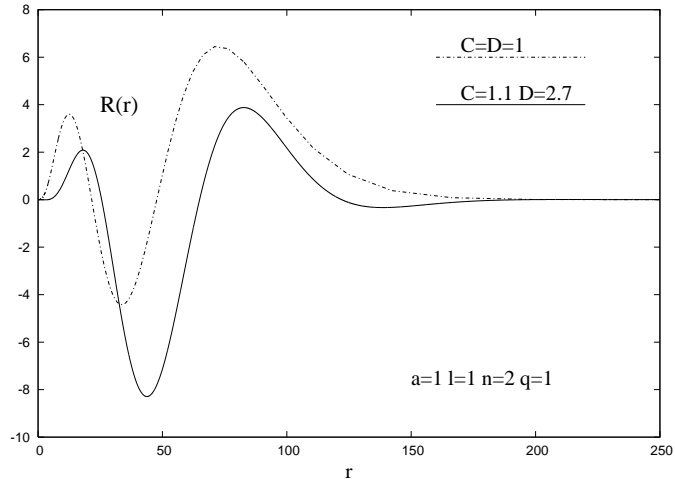} 
}}}
\caption{{\small The radial function $R(r)$ with a give set of quantum numbers
is plotted for a GPS
monopole background ($C=D=1$) and a IK metric.}}
\end{figure}
with $p_k$ real coefficients.
%
Supposing that this series has $n$ terms  and
replacing in (\ref{p1}),  one finds the quantization of the energy
\begin{eqnarray}
\label{p11}
E^2=\frac{2}{a^2}(C q^2-(1+n+l)^2+(1+n+l)\sqrt{(1+n+l)^2-2Cq^2+Dq^2})~,
\end{eqnarray}
with
$\lim_{n\to \infty} E= {2C q^2}/{a^2}$.
This result is found by taking the expression (\ref{f1}) for
 $K(r)$, or the large $r$ form (\ref{f2}).

Although the existence of the global solution (\ref{as3}), (\ref{series})
(with $E$ fixed by (\ref{p11}))
still requires an existence proof, this agrees with the closed form 
solution known for $C=D=1$ \cite{Cotaescu:1999gu}.
The expansion (\ref{series}) has only one
free coefficient,
which is fixed by the normalization condition (\ref{(scprod)}).
However, the coefficients $p_k$ satisfy a complicated recurence relation,
which we could not establish in the general case.

In practice, the solutions of the equation (\ref{(radeq)}) interpolating between
(\ref{as1}) and (\ref{as2})  can be constructed 
numerically  by using a standard ordinary
differential equation solver.
A plot of two typical radial functions with quantum numbers 
$l=1,~n=2,~q=1$ is given
in Figure 1 for $C=D=1$ (the case of a standard KK monopole background) and 
a IK geometry  with $C=1.1,~D=2.7$ (with  $a=1$ in both cases). 
These solutions decay exponentially and have a frequency fixed by (\ref{p11}).
The solution with an asymptotic oscillatory behaviour looks very similar 
to that
presented in Figure 2 for a metric background given by (\ref{bolt}).

One should also remark that, although
we have restricted  to a IK background because of their geometrical relevance,
a quantization relation similar to (\ref{p11}) is found for any regular 
 geometry with a
"nut" in the $t=const.$ section, provided that $E<|q|/(4n)$.

\subsection{A globally regular "bolt" background}
Although no closed form solution  is available in this case,
our results indicate that the properties of the
radial function $R(r)$ are rather similar to the case 
discussed above.
Here we shall restrict to the metric form  (\ref{bolt}), which is the only
case considered so far in the literature.
The radial function vanishes near  $r=r_b$, with 
\begin{eqnarray}
R(r)=(r-r_b)^{(1+4|q|)/2}c_0\left(1-\frac{2}{3m}(1-4|q|)(r-r_b)+\dots\right),
\end{eqnarray}
while the asymptotic form is still given by (\ref{asi1}), (\ref{asi2}). 

When taking a general solution on the form
$R(r) = (r-r_b)^{(1+4|q|)/2} e^{-\sqrt{ q^2/(4m^2)-E^2} r} P(r),$
and suppose a power series expansion for $P(r)$,
one finds that, for $E<q/2m$,
%
\begin{figure}[ht!]
\parbox{\textwidth}
{\centerline{
\mbox{
\epsfysize=18.0cm
\includegraphics[width=92mm,angle=0,keepaspectratio]{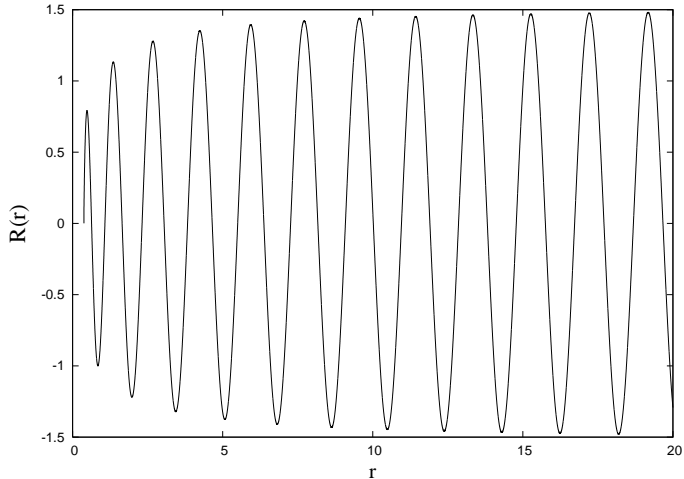} 
}}}
\caption{{\small 
The radial function $R(r)$  
is plotted for a $d=5$ metric which is the product of the $d=4$ Euclidean 
Taub-bolt solution 
with the real line. The parameters here are $m=1$, $l=1$, $q=0$ and $E=3$.}}
\end{figure}
%
the spectrum is quantized again according to
\begin{eqnarray}
E^2=\frac{q^2}{4m^2}-
\left( \frac{2n-1+4|q|}{5m}-\sqrt{\frac{(2n-1+4|q|)^2}{25m^2}-
\frac{q^2}{4m^2}}\right)^2~.
\end{eqnarray}
For $E>q/2m$, the spectrum is continuous, with an oscillatory behaviour 
of the radial wave function given by (\ref{asi2}).

The general solution is again constructed numerically and looks rather similar 
to the case of a IK background.
A typical oscillatory solution 
for a "bolt" background  with  $r_b=8/3$ is plotted in Figure 2.

\section{Solutions in a squashed Kaluza-Klein black hole background}

Apart from the geometries with a regular origin
considered in the previous section, the metric ansatz  (\ref{(met)})
presents another type of configurations. 
These are the squashed KK black holes, which enjoyed  
recently some 
interest, following the discovery by
Ishihara and Matsuno (IM) \cite{Ishihara:2005dp} of a 
 charged solution in the five dimensional Einstein-Maxwell theory with 
 some special properties. 
The horizon  of the IM black hole has $S^3$ topology, and
its spacelike infinity is a squashed sphere or $S^1$ bundle over $S^2$.
The mass and thermodynamics of this solution  have been discussed in 
\cite{Cai:2006td}. 
A vacuum black hole solution  with similar properties has been presented 
in \cite{Chen:1999rv}.
The IM black hole has been generalized in several directions, 
including 
KK rotating solutions with squashed horizon  \cite{Wang:2006nw},
an Einstein-Maxwell-dilaton version \cite{Yazadjiev:2006iv}
and solutions with nonabelian matter fields  \cite{Brihaye:2006ws}.

As mentioned above, the expression of the metric functions are fixed 
by the matter content
of the theory, the Einstein-Maxwell solution (\ref{IM}), 
(\ref{IM-U1}) being the most studied case
in the literature. 
However, in all cases, the squashed KK black holes
approach asymptotically a twisted
$S^1$ bundle over a four dimensional Minkowski spacetime, with the approximate
form of the metric functions
given by (\ref{inf-exp}).
For such solutions, the spacetime behaves as a five-dimensional black hole
near the horizon, while the dimensional reduction to four is realized 
in the far region.

The regular horizon is located at $r=r_h$, where the following expansion holds
(here we consider non-extremal solutions only)
\begin{eqnarray}
\label{eh-exp}
f(r)=f_h+\sum_{k\geq 1} f_{k }(r-r_h)^k,~g(r)=g_h+\sum_{k\geq 1} g_k(r-r_h)^k,
~h(r)= h_2(r-r_h)^2+\sum_{k\geq 3} h_k(r-r_h)^k,
\end{eqnarray}
where $f_h,~f_k,~g_h,~g_k, h_2,~h_k$
are a set of constants which are
fixed by the Einstein-matter field equations
(with $(f_h,~g_h,~h_2)>0$).
The Hawking temperature of the black holes, as computed from the 
surface gravity $T_H=\kappa/2\pi$
(with $\kappa^2=- \frac{1}{4}g^{tt}g^{rr}(\partial_r g_{tt})^2$),
is
$ T_H= \sqrt{ {h_2}/{f_h}}/(2\pi)$.

It is interesting to study the KG equation in this case and 
to see the differences with respect to the globally regular case.
The equation (\ref{(kg)}) is solved here with a suitable set of
boundary conditions at the inner $r=r_h$ boundary and at infinity. 
Some properties of the KG equation in a IM background have been 
discussed recently 
in Ref. \cite{Ishihara:2007ni}\footnote{Note that, however, a 
Schwarzschild-like coordinate system
is used in \cite{Ishihara:2007ni}.}.
Here we consider instead a generic line element, supposing 
the metric functions satisfy the asymptotic expansions (\ref{eh-exp}), 
(\ref{inf-exp}).
As expected, no closed form solution of the KG equation is available
in the black hole case.
Also, these solutions cannot be considered as a perturbation 
around a GPS background.

As usual with black holes \cite{Birrell:1982ix}, the general solution of the Schr\"odinger-type 
eq. (\ref{n2}) has the 
following form near the horizon of the black hole 
\begin{eqnarray}
\label{sol-eh1}
F=  \lambda_1 e^{-i E r_*}+ \lambda_2 e^{+i E r_*}~.
\end{eqnarray}
However, only those solutions are retained which represent no particles
coming out of the horizon of the black hole;
this is satisfied if we set
$\lambda_2=0$.
Expressing this result in terms of the radial function $R(r)$,
one finds the approximate form of the solutions as $r \to r_h$
\begin{eqnarray}
\label{sol-eh}
R(r)= \lambda_1\sqrt{r-r_h}~{\rm exp}\bigg(-i\frac{E}{2\pi T_H}\log(r-r_h)\bigg)
~(1+ c_1(r-r_h)+c_2(r-r_h)^2+\dots)~,
\end{eqnarray}
where the coefficients $c_k$ are fixed by the geometry
and the parameters $E$, $l$ and $q$, $e.g.$
one finds for the lowest order term
\begin{eqnarray}
\label{coeff-c1}
c_1=\frac{h_2\left(f_1g_hh_2r_h+f_h(4g_hh_2+g_1h_3r_h+g_hh_3r_h)
+4f_hg_h(f_hh_3-f_1h_2)r_h E^2\right)}
{4f_hg_hh_2r_h(h_2+2i\sqrt{f_h h_2}E)}~.
\end{eqnarray}
Considering now the solution in the large $r$ region,
one finds that the the asymptotic forms (\ref{asi1}), (\ref{asi2}) 
hold  in this case, too.
The $M/r$ coefficient which enters the expression of $g_{tt}$
 affects there only the next to leading 
order terms.
Different $e.g.$ from the Schwarzschild black hole case,
the spectrum of a KK black hole with squashed horizon 
contains an  exponentially decaying part 
for small enough frequencies ($E<|q|/2m$).
No quantization of frequencies occurs in the black hole case, however.
For any given background, the solution interpolating 
between (\ref{sol-eh}) and (\ref{asi1}), (\ref{asi2}) 
can be constructed numerically  by employing similar methods to those
used in the globally regular case.

We close this Section by remarking that the Ref. \cite{Ishihara:2007ni} 
presented arguments that the
luminosity of the Hawking radiation of the Einstein-Maxwell  black hole 
solution (\ref{IM}) encodes 
information about the size of the extra-dimension.
This result is found by computing the absorbtion probability 
by matching the asymptotic expansions of the solutions of the KG equation
and is
likely to hold for any KK squashed black hole.

\section{Further remarks}
A considerable number of attempts have been carried out 
in order to solve the matter field equations in a four dimensional curved 
background.
However, the corresponding task for higher dimensional 
solutions is still in its early stages, 
especially for spacetimes with a more complicated topological structure.

In this work we have analyzed the basic properties of the scalar
wave equation in a generalized 
KK monopole spacetime.
Our results are relevant when considering the scalar field quantization
in these metric backgrounds.
As a characteristic feature, we noticed the 
existence of an exponentially damped part of the spectrum, 
for small enough frequencies.
In the globally regular case, the frequencies are quantized, the 
properties of the KG equation in a GPS monopole background
being generic in this case. 
For large enough frequencies, a continuum wave spectrum is found instead.
In the well-known GPS case, these results are in agreement with the 
behaviour of classical test particles.
It would be interesting to consider the geodesic motion
 in a general background (\ref{(met)}).

Similar solutions can be constructed for higher spin fields,
the case of Dirac equation in a general metric background (\ref{(met)})
being currently under study. 

It is interesting to compare the properties of the scalar field we have 
found for a background (\ref{(met)})  
approaching a twisted
$S^1$ bundle over a four dimensional Minkowski spacetime 
with  the case of a topologically trivial KK background.
Such geometries are described by a metric ansatz\footnote{The functions 
$f,~g,~h$
here have nothing to do with those of the generalized KK configurations.}
\begin{eqnarray}
\label{(met-KK)}
ds^{2}=-h(r) dt^{2}+f(r) \big(dr^{2}+ r^{2}(d\theta^{2}+ 
\sin^{2}\theta\, d\phi^{2})\big)+ g(r) d\chi ^{2}\,,  
\end{eqnarray}
and approach at infinity the four dimensional Minkowski spacetime times 
a circle (here $0\leq \chi \leq L$, with $L$ arbitrary).
The simplest solution   of this type 
is found by trivially extending to five dimensions  
the Schwarzschild black hole solution  to vacuum
Einstein equations in four dimensions,
and  corresponds to a  uniform black string  
with horizon topology $S^2\times S^1$.
(Note that nontrivial globally regular solutions are found within the 
metric ansatz (\ref{(met-KK)})
only in the presence of nonlinear matter fields, see $e.g.$ 
the solutions in \cite{Volkov:2001tb}).
 Separating the variables in the KG equation (\ref{(kg)}) 
by writing (with $\rho$ still given by (\ref{ro}))
\begin{equation}
\psi(x)=\frac{1}{\rho(r)} 
R_{E,l}^q (r)\,Y_{l,m} (\theta,\phi )e^{i(q\chi -E t)}~,
\end{equation}
where $Y_{l,m} (\theta,\phi )$
are the spherical harmonics and $q=2n\pi/L$  (with $n$ an integer), 
one obtains the radial equation for $R$ in the form
\begin{eqnarray}
\label{(radeq2)}
&&\left[-\frac{d^2}{dr^2}+\frac{l(l+1)}{r^{2}} +
\frac{f}{g} q^{2}+\frac{1}{\rho}\frac{d^{2}\rho}{
dr^2}\right] R_{E,l}^q (r)
= E^2\frac{f}{h}R_{E,l}^q(r)~.
\end{eqnarray}
Comparing with (\ref{(radeq)}) reveals that the main effect at this level of a
KK monopole asymptotic structure of the metric background 
 is to add a new term to the potential in the
Schr\"odinger-like equation satisfied by the radial wave function. This term 
decreases the height of
the potential and becomes relevant for high energies.
Note also the existence in this case, too, of 
an exponentially decaying part of the spectrum for
$E<|q|$.

For the case of a vacuum black string,
the solutions of the equation (\ref{(radeq2)}) are found by taking 
$E\to \sqrt{E^2-q^2}$
in the solutions corresponding to a $d=4$ Schwarzschild black hole background.

\subsection*{Acknowledgments}
We thank  Cristian Stelea for  valuable remarks on a
draft of this paper.
The work of E. R. was carried out in the framework of Enterprise--Ireland Basic Science
Research  Project SC/2003/390 of Enterprise-Ireland.  
The work of M. V.  was supported in part by the CNCSIS-Romania grant 1154/2007 
and CEEX-Romania program CEx 05-D11-49.

%
\end{document}